\documentclass[a4paper,11pt]{article}
\usepackage{amsmath}
\usepackage{bm}
\usepackage{graphicx}
\usepackage{comment}
\begin{document}

\title{A model of dense-plasma atomic structure for equation-of-state calculations}

\author{J C Pain\footnote{Commissariat \`a l'Energie Atomique, CEA/DIF, B.P. 12, 91680 Bruy\`{e}res-le-Ch\^{a}tel Cedex, France, jean-christophe.pain@cea.fr}}

\maketitle

\begin{abstract}
A model of dense plasmas relying on the superconfiguration approximation is presented. In each superconfiguration the nucleus is totally screened by the electrons in a Wigner-Seitz sphere (ion-sphere model). Superconfigurations of the same charge are grouped into ions. It is shown that boundary values of the wavefunctions play a crucial role in the form of the Virial theorem from which the pressure formula is derived. Finally, a condition is presented and discussed, which makes the ion-sphere model variational when bound electrons are treated quantum-mechanically and free electrons quasi-classically.
\end{abstract}

\section{Introduction}

Theoretical studies of electronic structure of high-energy-density plasmas are of great interest for the understanding of radiative transfer and equation of state (EOS) and for the simulation of laser-driven experiments. Such investigations play a major role in astrophysics and inertial confinement fusion. The present work consists in a theoretical study of atomic physics of ions in plasmas taking screening effects into account. It is essential, in order to describe dense plasmas and especially their ability to absorb electromagnetic radiations, to characterize ions in the most realistic way, through a consistent thermodynamic modeling. 

Photo-absorption calculations using self-consistent ``Detailed-Term Accounting'' (DTA) or ``Detailed-Configuration Accounting'' (DCA) have been performed (see for instance Goldberg \emph{et al.} \cite{gol} or Rozsnyai \emph{et al.} \cite{roz0}). In this framework, all ionic species are treated in all their physically realistic configurations. However, the number of relevant configurations in a plasma can be really huge, especially for high values of atomic number $Z$, and all of them can not be taken into account. Indeed, requiring that each configuration has its own basis of one-electron wavefunctions and including configuration mixing leads to a high numerical cost. The superconfiguration method has been invented \cite{bar0} in order to overcome this problem in an approximate way and to provide more accurate results than Average-Atom models. This approach includes relaxation effects (different basis of wavefunctions for the upper and lower states involved in a transition), and allows inclusion of configuration interaction between relativistic sub-configurations of a non-relativistic configuration \cite{bar1, bar2, bar3}.

As a configuration is defined by shells occupied by an integer number of bound electrons, a superconfiguration (SC) is defined by supershells, which are groups of shells close in energy \cite{bar0} (\textit{i.e.} whose energies differ from less than $k_BT$). A SC can be represented by

\begin{equation}
\Xi=\underbrace{(\alpha_1\alpha_2\cdots\alpha_{N_1})^{a_1}}_{\sigma_1}\underbrace{(\beta_1\beta_2\cdots\beta_{N_2})^{a_2}}_{\sigma_2}\underbrace{(\gamma_1\gamma_2\cdots\gamma_{N_3})^{a_3}}_{\sigma_3}\cdots,
\end{equation}
 
where $\{\alpha_i$, i=1, $N_1\}$ are the orbitals of supershell $\sigma_1$, $\{\beta_i$, i=1, $N_2\}$ are the orbitals of supershell $\sigma_2$ and $\{\gamma_i$, i=1, $N_3\}$ are the orbitals of supershell $\sigma_3$. $a_1$ is the number of electrons in supershell $\sigma_1$, $a_2$ the number of electrons in supershell $\sigma_2$, \textit{etc.} The Super Transition Array (STA) formalism developed by Bar Shalom \emph{et al.} \cite{bar0} enables one to calculate photo-absorption spectra within the SC approximation. In our model, the nucleus is totally screened by electrons in a sphere (ion-sphere model) which is usually named, as referred to solid-state physics, a Wigner-Seitz (WS) sphere, and which radius is given by

\begin{equation}
r_{ws}=\left(\frac{3}{4\pi n_i}\right)^{1/3},
\end{equation} 

where $n_i$ represents ionic density (number of ions per volume unit). The ionic coupling parameter of the plasma is defined as the ratio of ionic Coulomb potential energy and thermal kinetic energy 

\begin{equation}
\Gamma_{ii}=\frac{Z_{eff}^2}{r_{ws}T},
\end{equation} 

where $Z_{eff}$ is the effective (average) charge of the plasma. Equation of state of dense matter can be addressed by several techniques such as, for instance, chemical-picture models, quantum-molecular dynamics, path-integral Monte Carlo, or average-atom models. The latest are particularly well suited for high-Z elements, and hot dense plasmas with coupling parameter $1\leq\Gamma_{ii}\leq 10$, namely strongly correlated plasmas. The model presented here is close to AA models, but constitutes an attempt to include better (less averaged) atomic physics. For each SC, bound electrons are treated quantum mechanically and free electrons within the Thomas-Fermi (TF) approximation (hybrid model). Once their electronic structure has been calculated, SCs of the same charge can be grouped into ions. At this stage, the plasma can be considered as an ensemble of ions, containing bound electrons, and immersed in a free-electron gas. It was shown \cite{pai, pai2} that, provided that ions of different charges have different volumes, the electronic pressure, calculated together with ionic volumes in a self-consistent way, must be equal for each ion. The improvement in thermodynamic consistency brought by the diversity of ionic volumes (in other words the ``redefinition'' of ions) encourage us to apply this approach to EOS calculations.

The purpose of the work presented in \cite{pai2} was to propose a method for the treatment of mixtures, and the issue of thermodynamic consistency was not addressed. In the present work, we focus on the expression of pressure, and especially on the proper form of the quantum Virial theorem. It is shown that the universal form of the Virial theorem for a quantum-statistical model (\textit{i.e.} a Virial theorem that does not depend on the boundary conditions of the wavefunctions) is different from the one proposed in \cite{per,leg}, which relies in an intrinsic way on Neumann or Dirichlet conditions. Moreover, we find that the pressure obtained from the generalized Virial theorem presented here is equal to the stress-tensor formula proposed in \cite{mor,ble3} and can be obtained by minimization of the free energy. Such a formula can be also applied to free-electrons \cite{pai3}, but for a sake of simplicity we consider here the ``hybrid'' model, in which free electrons are treated semi-classically. 

The transformation of an electron state from being bound to being a part of the continuum is gradual \cite{koh}. However, although the hybrid model provides an accurate description of strongly correlated plasmas, the confinement of an ion in a finite volume may still lead, in our model and when the matter density is sufficiently high, to unacceptable discontinuities in the thermodynamic functions when a bound level disappears (pressure ionization). Such discontinuities in thermodynamic functions are due to the fact that it is difficult to obtain a variational formulation of the problem.

Our EOS model relying on a self-consistent-field calculation of screened SCs with TF free electrons and quantum bound electrons is explained in section \ref{sel}. It is shown in section \ref{tow} that boundary values of the wavefunctions play a crucial role in the continuity of pressure and in the expression of the Virial theorem from which the pressure formula is derived. In the same section, a condition is proposed which should make variational the ion-sphere model when bound electrons are treated quantum-mechanically and free electrons quasi-classically. Its impact on the expression of electronic pressure within the SC approximation is discussed.

\section{\label{sel}Equation of state relying on a self-consistent calculation of screened SCs}

\subsection{State of the art}

Blenski \emph{et al.} \cite{ble1,ble2} have developed a model of matter based on a self-consistent-field (SCF) calculation of SCs. The first step is an Average-Atom (AA) calculation \cite{roz}, in order to get the list of relevant orbitals necessary for the definition of supershells and SCs. Each SC is a grand canonical ensemble of bound electrons. In local thermodynamic equilibrium, \textit{i.e.} when collisions between particles (electrons and ions) are sufficiently frequent (when the density is sufficiently high) and when the mean free path is small compared to the characteristic length of the system, the plasma is thermalized and the probability $W_{\Xi}$ of a given SC $\Xi$ is given by Boltzmann's law

\begin{equation}\label{bolf}
W_{\Xi}\propto\exp [-\frac{\Omega_{\Xi}}{T}], 
\end{equation}

where $T$ is temperature and $\Omega_{\Xi}$ the grand potential of the SC. When the matter density increases, the de Broglie length becomes of the same order as the inter-electronic distance. Therefore one has to take quantum effects into account in the electron-electron interaction. The number of bound electrons in a supershell $\sigma$ is defined as

\begin{equation}
N_{\sigma}=\sum_{k\in\sigma}g_kf_k,
\end{equation}

$g_k$ and $f_k$ being respectively the degeneracy and the occupation factor of sub-shell $k$. The number of bound electrons in a SC $\Xi$ is then given by

\begin{equation}\label{inte}
N_{\Xi, b}=\sum_{\sigma\in\Xi}N_{\sigma}.
\end{equation}

In the present work, occupations factors are evaluated using

\begin{equation}\label{expf1}
f_k=\frac{1}{1+e^{(\epsilon_k-\mu_{\sigma})/T}},
\end{equation}

$\mu_{\sigma}$ being adjusted so that 
the $N_{\sigma}$ is an integer \cite{ble2}. However, we also have the possibility to evaluate $f_k$ using the exact average value calculated with the partition functions, according to \cite{bar0}:

\begin{equation}\label{expf2}
f_k=\frac{1}{1+\frac{M_{Q_{\sigma}}^{(k)}}{M_{Q_{\sigma}-1}^{(k)}}e^{\epsilon_k/T}},
\end{equation}

where $M_{Q_{\sigma}}^{(k)}$ is the partition function of supershell $\sigma$ containing $Q_{\sigma}$ electrons, in which the degeneracy of orbital $k$ has been reduced by one. It was shown in \cite{gil} that expressions (\ref{expf1}) and (\ref{expf2}) can be different. If $f_k$ is defined according to equation (\ref{expf1}), the model is a superposition of a number of AA models (one AA model for each charge state). Definition (\ref{expf2}) is more consistent with the fundamentals of the SC approach. The free-elecron gas is considered as a partially degenerate ideal gas of non relativistic spin $1/2$ particles in the electrostatic potential $v(r)$. The total number of free electrons of SC $\Xi$ can be written, within the TF \cite{tho,fer,kir,fey} model

\begin{equation}
N_{\Xi,f}=\frac{2}{(2\pi)^3}\int_{V_{\Xi}}d^3r\int_{\sqrt{-2v(r)}}^{\infty}4\pi p^2dp\frac{1}{1+\exp[\frac{p^2/2+v(r)-\mu_{\Xi}}{T}]},
\end{equation}

where $V_{\Xi}=\frac{4}{3}\pi r_{\Xi}^3$ is the volume associated to SC $\Xi$, $r_{\Xi}$ being the Wigner-Seitz radius associated to SC $\Xi$ and $\mu_{\Xi}$ the grand canonical chemical potential associated to the preservation of the total number of electrons

\begin{equation}
N_{\Xi, b}+N_{\Xi,f}=Z,
\end{equation}

$Z$ representing atomic number. The main advantage of TF approximation is that it does not include any linear response approximation. In other words, it is not necessary to assume that the ion-electron interaction is weak compared to electronic kinetic energy. Nevertheless it is important to remind that the central-field approximation requires a large number of electrons, \textit{i.e.} elements with a high atomic number. The SCF calculation of a SC provides the potential, a one-electron basis and chemical potentials (which define occupation factors) for the supershells and for the free-electron gas. The energy of a configuration is written as follows \cite{ble1}

\begin{equation}
E_c=\sum_kg_kf_kI_k+2E_c^{(bb)}+E_c^{(bf+f)},
\end{equation}

where $E_c^{(bb)}$ represents interaction energy between bound electrons:

\begin{equation}
E_c^{(bb)}=\frac{1}{2}\sum_{k,k'}g_kf_k(g_{k'}f_{k'}-\delta_{k,k'})V_{kk'}.
\end{equation}

Superscript $(bf)$ denotes interactions between bound and free electrons and $(f)$ the contribution of free electrons. The quantity $I_k$ stands for the one-particle operator integral:

\begin{equation}
I_k=\epsilon_k-\int [V(r)+\frac{Z}{r}]]y_k^2(r)dr
\end{equation}

where $\epsilon_k$ represents the energy of orbital $k$ and $y_k$ the radial part of the corresponding wavefunction multiplied by $r$. The energy of a configuration $c$ belonging to SC $\Xi$ is replaced by

\begin{equation}
E_c\approx\sum_kg_kf_kI_k+<2E_c^{(bb)}+E_c^{(bf+f)}>_{\Xi}
\end{equation}

in the Boltzmann factors. 
The contribution $2E_c^{(bb)}+E_c^{(bf+f)}$ being averaged over all the configurations in SC $\Xi$, one can define a partition function $M_{\Xi}$ allowing one to carry out calculations as in the independent-particle approximation

\begin{equation}
M_{\Xi}=\sum_{c\in\Xi}g_c\exp[-\frac{E_c}{T}],
\end{equation}

where $g_c$ represents the total degeneracy of a given configuration. The grand potential of SC $\Xi$ reads

\begin{equation}
\Omega_{\Xi}=F_{\Xi}-\mu_{\Xi}Z=-T\ln[M_{\Xi}]-TS_{\Xi}^{(f)}+F_{\Xi,xc}-\mu_{\Xi}Z,
\end{equation}

where $F_{\Xi,xc}$ is the exchange-correlation contribution at finite temperature, calculated using the formulas of Iyetomi and Ichimaru \cite{iye} within the local density approximation (LDA). These expressions result from calculations of the free-energy for the One-Component Plasma (OCP) in the HyperNetted Chain (HNC) approximation. The term $S_{\Xi}^{(f)}$ represents the entropy of free electrons.

The fact that the quadratic terms are averaged over all the configurations of a given SC constitutes the main approximation of the method. The effect of such correlations has been studied in \cite{fau} in the context of the AA screened-hydrogenic model. In general, the ionic distribution with correlations has smaller wings and a higher maximum (\textit{i.e. } a higher kurtosis). 

\subsection{\label{iso}Isobaric approach}

Preservation of total volume ($\sum_{\Xi}W_{\Xi}V_{\Xi}=V$ associated to Lagrange multiplier $P$) and normalization of SC probabilities ($\sum_{\Xi}W_{\Xi}=1$ associated to Lagrange multiplier $B$) have to be ensured, as well as normalization of bound electrons wavefunctions, \textit{i.e.} $\int_{V_{\Xi}}\psi_k^*(\vec{r})\psi_k(\vec{r})d^3r=1, \;\;\forall k\in\sigma\in\Xi$, associated to Lagrange multiplier $\lambda_k$. Furthermore, the fact that each supershell has an integer number of bound electrons $N_{\sigma}$ (see equation (\ref{inte})) implies the introduction of Lagrange multiplier $\mu_{\sigma}$. The constrainted grand potential thus reads

\begin{eqnarray}\label{granpot}
\tilde{\Omega}&=&\sum_{\Xi}W_{\Xi}\Omega_{\Xi}+B[\sum_{\Xi}W_{\Xi}-1]+P[\sum_{\Xi}W_{\Xi}V_{\Xi}-V]+T\sum_{\Xi}W_{\Xi}\ln W_{\Xi}\nonumber \\
&&+\sum_{\Xi}W_{\Xi}\sum_{\sigma\in\Xi}\sum_{k\in\sigma}\lambda_k[\int_{V_{\Xi}}\psi_k^*(\vec{r})\psi_k(\vec{r})d^3r-1]\nonumber \\
&&+\sum_{\Xi}W_{\Xi}\sum_{\sigma\in\Xi}\mu_{\sigma}[\sum_{k\in\sigma}g_kf_k-N_{\sigma}]\nonumber \\
&=&\sum_{\Xi}W_{\Xi}\tilde{\Omega}_{\Xi}+B[\sum_{\Xi}W_{\Xi}-1]+P[\sum_{\Xi}W_{\Xi}V_{\Xi}-V]+T\sum_{\Xi}W_{\Xi}\ln W_{\Xi}, 
\end{eqnarray}

where

\begin{equation}
\tilde{\Omega}_{\Xi}=\Omega_{\Xi}+\sum_{\sigma\in\Xi}\sum_{k\in\sigma}\lambda_k[\int_{V_{\Xi}}\psi_k^*(\vec{r})\psi_k(\vec{r})d^3r-1]+\sum_{\sigma\in\Xi}\mu_{\sigma}[\sum_{k\in\sigma}g_kf_k-N_{\sigma}].
\end{equation}

Variations with respect to volume $V_{\Xi}$ and probability $W_{\Xi}$ give respectively 

\begin{equation}\label{press1}
P_{\Xi}\equiv-\frac{\partial\tilde{\Omega}_{\Xi}}{\partial V_{\Xi}}\Bigl|_{T}=P \;\; ; \;\; W_{\Xi}\propto\exp [-\frac{\tilde{\Omega}_{\Xi}+PV_{\Xi}}{T}],
\end{equation}

which means \cite{pai} that all SCs have the same pressure and that it is necessary to include a work term $PV_{\Xi}$ in the Boltzmann factor which plays a crucial role \cite{pai2}. All pressures are calculated in a self-consistent way and equalized by a multi-dimensional Newton-Raphson method.

\begin{figure}
\begin{center}
\includegraphics[width=8.6cm, trim=0 0 0 -60]{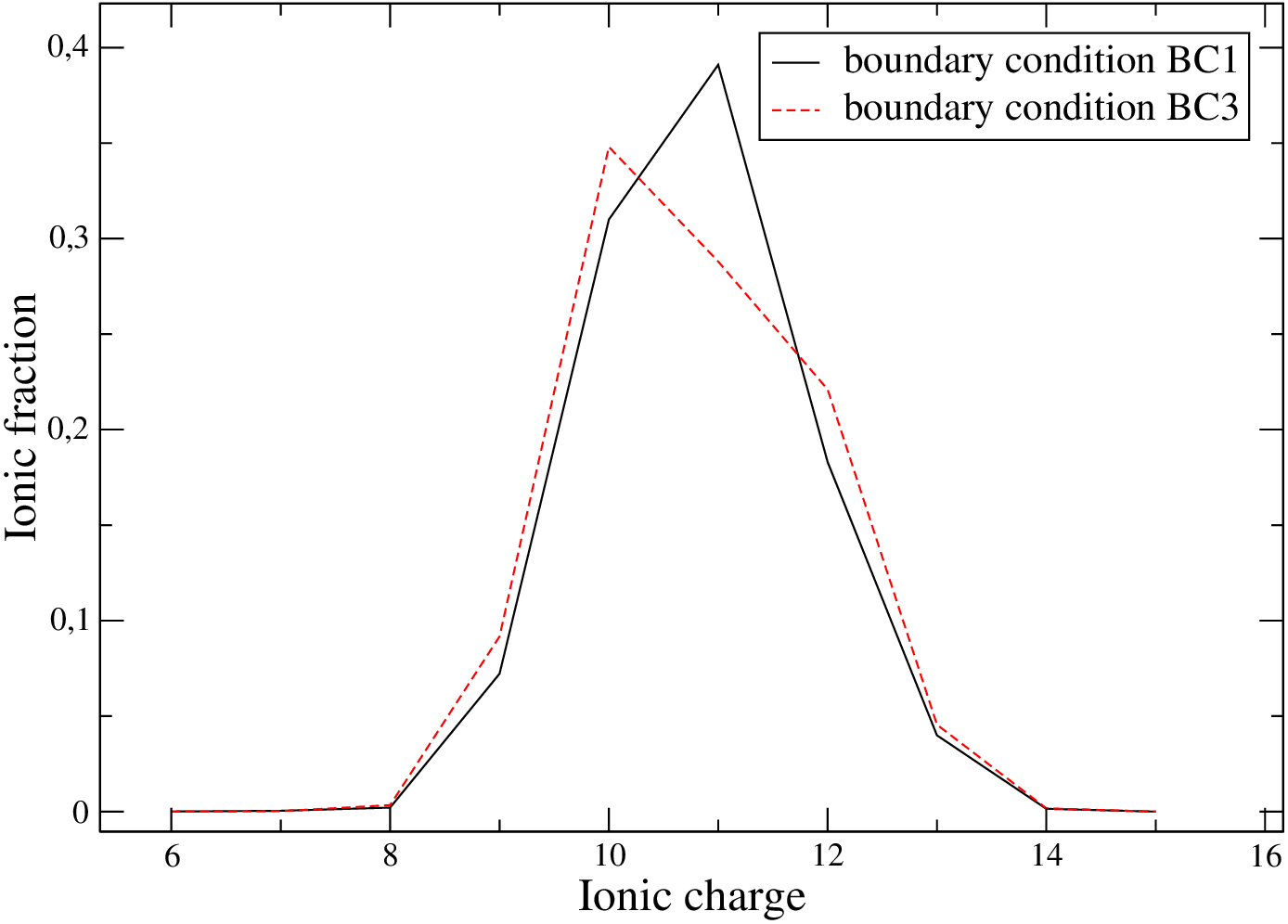}
\end{center}
\caption{\label{fig1}Ionic distribution for a vanadium plasma (V, Z=23) at T=100 eV and $\rho$=2.5 g/cm$^3$ with boundary conditions BC1 and BC3.}
\end{figure}

\subsection{Calculation of pressure}

The TF free-electron pressure dominates highly bound-electron and exchange-correlation pressures. The behaviour of the free-electron pressure

\begin{equation}
P_{\Xi,f}(r_{\Xi})=\frac{2}{3(2\pi)^3}\int_{\sqrt{-2v(r_{\Xi})}}^{\infty}4\pi p^4dp\frac{1}{1+\exp[\frac{p^2/2+v(r_{\Xi})-\mu_{\Xi}}{T}]}
\end{equation}

can be understood by the following quantity 

\begin{equation}
R=\frac{\int_{\sqrt{-2v(r_{\Xi})}}^{\infty} p^4dp\frac{1}{1+\exp[\frac{p^2/2+v(r_{\Xi})-\mu_{\Xi}}{T}]}}{\int_{0}^{\infty} p^4dp\frac{1}{1+\exp[\frac{p^2/2+v(r_{\Xi})-\mu_{\Xi}}{T}]}}=\frac{A(r_{\Xi})}{B(r_{\Xi})},
\end{equation}

ratio of pressure with free electrons described in the TF approximation and pressure with all electrons described in the TF approximation. When the density is low, 

\begin{equation}
A(r_{\Xi})\approx \frac{1}{T^{3/2}}\int_0^{\infty}[2(x-v(r_{\Xi}))]^{1/2}e^{-\frac{x}{T} }dx,
\end{equation}

and

\begin{equation}\label{equa}
B(r_{\Xi})\approx \frac{e^{-v(r_{\Xi})/T}}{T^{3/2}}\int_0^{\infty}[2(x-v(r_{\Xi}))]^{1/2}e^{-\frac{x}{T} }dx. 
\end{equation}

Taking into account the fact that the potential $v$ behaves like $\frac{A}{r_{\Xi}}$, $A$ being a non-dimensional constant, equation (\ref{equa}) leads to

\begin{equation}
R\approx\frac{e^{-v(r_{\Xi})/T}B(r_{\Xi})}{B(r_{\Xi})}
\approx \exp[-\frac{c}{Tr_{\Xi}}],
\end{equation}

where $c$ is a positive constant. Hence, when matter density increases, ratio $R$ decreases. Three-body recombination is the dominant process. At high density, kinetic energy $E_{kin}$ behaves like $r_{\Xi}^{-2}$ and potential energy $E_{pot}$ like $r_{\Xi}^{-1}$. The Virial theorem \cite{fey} $3PV_{\Xi}=2E_{kin}+E_{pot}$ indicates that, when density increases, kinetic energy becomes larger than potential energy and total pressure is systematically positive and increases. But it is not physically acceptable that free-electron pressure remains always positive. That is the reason why one has to take exchange-correlation phenomena at finite temperature into account. The exchange-correlation pressure is given by \cite{leg}:

\begin{eqnarray}
P_{\Xi,xc}&=&n_{\Xi}\frac{\partial f_{\Xi,xc}}{\partial n_{\Xi}}\Big|_{r_{\Xi}}-f_{\Xi,xc}(r_{\Xi})\nonumber \\
&=&-n_{\Xi}(r_{\Xi})v_{xc}(r_{\Xi})+\frac{1}{3V_{\Xi}}\int_0^{r_{\Xi}}v_{xc}(r)\frac{d(r^3n_{\Xi}(r))}{dr}4\pi dr,
\end{eqnarray}

where $F_{\Xi,xc}=\int_{V_{\Xi}} f_{\Xi,xc}d^3r$. Bound-electron pressure is calculated via a stress-tensor formula proposed by More \cite{mor} 

\begin{equation}\label{expresso}
P_{\Xi,b}=\frac{1}{4\pi r_{\Xi}^2}\sum_{n,l}(2l+1)f_{nl}\Big\{\Big[\frac{dR_{nl}}{dr}(r_{\Xi})\Big]^2-R_{nl}(r_{\Xi})\frac{d^2R_{nl}}{dr^2}(r_{\Xi})\Big\},
\end{equation}

where $R_{nl}$, radial part of the wavefunction $\psi(\vec{r})=R_{nl}(r)Y_{lm}(\theta,\phi)\chi_s$, is solution of the Schr\"odinger equation

\begin{equation}
-\frac{1}{2}\frac{\partial^2}{\partial r^2}[rR_{nl}(r)]-[\epsilon_{nl}-\frac{l(l+1)}{2r^2}-v(r)]rR_{nl}(r)=0
\end{equation}

and $f_{nl}$ is the Fermi-Dirac (FD) occupation factor of $nl$ orbital belonging to the average configuration representative of SC $\Xi$. Introducing the logarithmic derivative $D_{nl}=\frac{r}{R_{nl}(r)}\frac{d R_{nl}(r)}{dr}\Big|_{r_{\Xi}}$, equation (\ref{expresso}) can be written

\begin{eqnarray}\label{expdnl}
P_{\Xi,b}&=&\frac{1}{4\pi}\sum_{n,l}(2l+1)f_{nl}[R_{nl}(r_{\Xi})]^2\times\nonumber \\
&&\Big[\frac{D_{nl}^2+2D_{nl}-l(l+1)}{r_{\Xi}^2}+2(\epsilon_{nl}-v(r_{\Xi}))\Big].
\end{eqnarray}

Let us consider three particular boundary conditions \cite{roz}

\begin{eqnarray}
R_{nl}(r_{\Xi})\propto\exp[-\sqrt{-2\epsilon_{nl}}r_{\Xi}] \label{boco1}\\
\frac{1}{R_{nl}}\frac{dR_{nl}}{dr}(r_{\Xi})=0 \label{boco2}\\
R_{nl}(r_{\Xi})=0. \label{boco3}
\end{eqnarray}

Condition (\ref{boco1}), named BC1 in the following, is linked to the fact that outside the ion sphere, wavefunctions are given by Bessel functions (see section \ref{cha}). In solid-state physics, condition (\ref{boco2}), named BC2 in the following, is a periodicity condition, associated with the fact that all polyhedral Wigner-Seitz cells are identical. But in the new constant-pressure approach, all ion spheres have different volumes, which makes condition BC2 irrelevant. It is worth mentioning that BC2 and BC3 predict two different eigenenergies, the lowest being provided by BC2. These two values can be regarded as the lower and upper limits of an energy band \cite{roz}. Bands can give a better treatment of pressure ionization (see section \ref{thr}). However, a band must be populated by a given density of states, which is unknown. Condition (\ref{boco3}), named BC3 in the following, gives systematically a positive bound-electron pressure (see equation (\ref{expresso})), on the contrary to condition BC2, which leads to a negative contribution. If condition BC1 is retained, pressure can be positive or negative. Total pressure is then given by $P_{\Xi}=P_{\Xi,f}+P_{\Xi,b}+P_{\Xi,xc}$. Table \ref{tab1} displays the values of free-electron pressure, bound-electron pressure, exchange-correlation pressure and total pressure for a vanadium plasma at $\rho$=25 g/cm$^3$ and temperatures 1 eV and 10 eV for boundary condition BC1. Free-electron pressure gives the highest contribution. Table \ref{tab2} presents, for boundary conditions BC1, BC2 and BC3, pressure calculated from an AA calculation and pressure calculated from a SC calculation in the case of a vanadium plasma at T=100 eV and $\rho$=2.5 g/cm$^3$. For each boundary condition, the differences between AA and SC results are quite small, but the values from a SC calculation should be more accurate, since they take into account the multiplicity and degeneracy of electronic configurations, overcoming in that way the ``rigidity'' of the AA model. SCs of the same charge can be grouped into ions. In both cases, (AA or SC calculations), values from different boundary conditions are quite close. Table \ref{tab2bis} presents the values of exchange-correlation pressure in the same conditions as Table \ref{tab2}. Exchange-correlation pressure is always negative, Condition BC3 provides the lowest value and in that case the contrihution of exchange-correlation to total pressure is very small. It is worth noticing that in the SC calculation, boundary conditions BC2 and BC3 give the same value, which is not the case in the AA calculation. Table \ref{tab3} contains the values of partial pressure of each ion charge in a vanadium plasma at T=100 eV and $\rho$=2.5 g/cm$^3$. Table \ref{tab4} contains the values of partial densities of the charge states in the isobaric approach. In that case, all the ions have the same boundary pressure and a different partial density. Calculations have been performed with boundary condition BC1.

\begin{figure}
\begin{center}
\includegraphics[width=8.6cm, trim=0 0 0 -60]{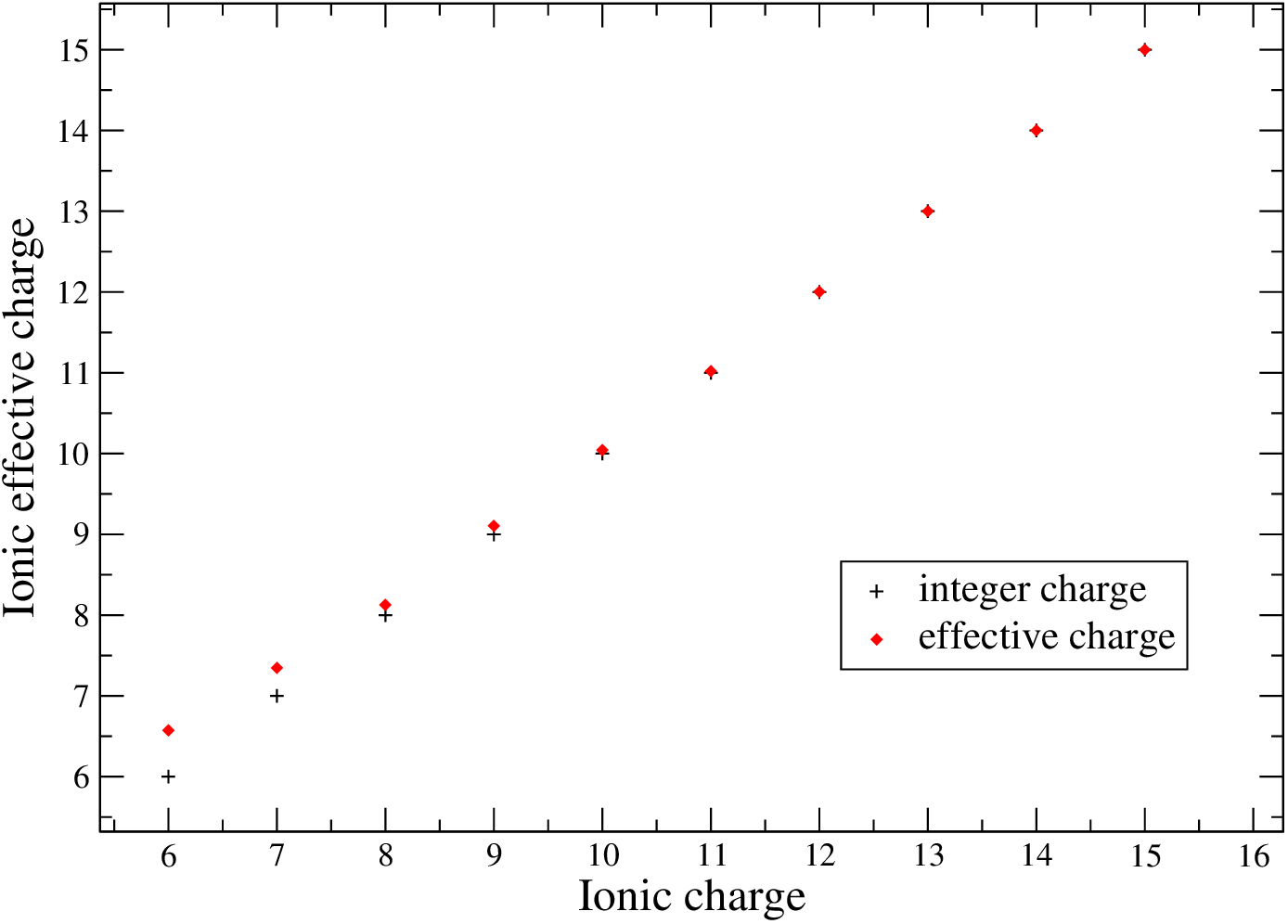}
\end{center}
\caption{\label{fig2}Real ionic charge for a vanadium plasma (V, Z=23) versus integer charge for ions $V^{5+}$ to $V^{13+}$ at T= 100 eV and $\rho$=2.5 g/cm$^3$.}
\end{figure}

\begin{table}
\caption{\label{tab1}Total pressure and exchange-correlation pressure for two values of matter temperature (1 eV and 10 eV) in the case of a vanadium plasma at $\rho$=25 g/cm$^3$ for boundary condition BC1.}
\begin{tabular}{|c|c|c|c|c|}\hline
Temperature (eV) & $P_{f}$ (Mbar) & $P_{b}$ (Mbar) & $P_{xc}$ (Mbar) & $P_{tot}$ (Mbar)\\\hline\hline
1 & 83.93 & -2.91 & -17.06 & 66.57 \\\hline
10 & 94.20 & -2.87 & -17.93 & 75.97 \\\hline
\end{tabular}
\end{table}

\begin{table}
\caption{\label{tab2}Total pressure evaluated from an AA model and a SC calculation for boundary conditions BC1, BC2 and BC3 for an vanadium plasma at T=100 eV and $\rho$=2.5 g/cm$^3$.}
\begin{tabular}{|c|c|c|}\hline
Boundary condition & $P_{tot}$ AA (Mbar) & $P_{tot}$ SC (Mbar)\\\hline\hline
BC1 & 42.2310 & 43.6674\\\hline
BC2 & 43.6943 & 43.8372\\\hline
BC3 & 43.6931 & 43.8362\\\hline
\end{tabular}
\end{table}

\begin{table}
\caption{\label{tab2bis}Exchange-correlation pressure evaluated from an AA model and a SC calculation for boundary conditions BC1, BC2 and BC3 for an vanadium plasma at T=100 eV and $\rho$=2.5 g/cm$^3$.}
\begin{tabular}{|c|c|c|}\hline
Boundary condition & $P_{xc}$ AA (Mbar) & $P_{xc}$ SC (Mbar)\\\hline\hline
BC1 & -0.7653 & -0.7971\\\hline
BC2 & -0.7813 & -0.8003\\\hline
BC3 & -0.7968 & -0.8003\\\hline
\end{tabular}
\end{table}

\subsection{\label{cha}Characterization of ionic species}

The value of quantity $L$ (pressure, \textit{etc.}) for ion $i$ is given by

\begin{equation}
L=\sum_{\Xi\in i}W_{\Xi}L_{\Xi}.
\end{equation}

Figure \ref{fig1} shows the ionic distribution for boundary conditions BC1 and BC2. The relative fraction of the ions depends strongly on the boundary condition. Some bound-electron wavefunctions can take significant values at the boundary of the WS sphere. When that happens, \textit{i.e.} when the wavefunction is non zero at the boundary, one can consider that the wavefunction extends outside the sphere. In other words, a corresponding bound state has a non-zero probability to be outside the sphere, where the potential is zero. In this region, radial part $R_{nl}$ of the bound-electron wavefunction is solution of the Schr\"odinger equation

\begin{equation}
-\frac{1}{2}\frac{\partial^2}{\partial r^2}[rR_{nl}(r)]-\Big[\epsilon_{nl}-\frac{l(l+1)}{2r^2}\Big]rR_{nl}(r)=0
\end{equation}

and $R_{nl}$ can be expressed by a Bessel function of the third kind

\begin{equation}
R_{nl}(r)=\sqrt{-\frac{\pi\epsilon_{nl}}{2r}}K_{l+\frac{1}{2}}(\sqrt{-\epsilon_{nl}}r)\equiv \sqrt{-\epsilon_{nl}}f_l(\sqrt{-\epsilon_{nl}}r).
\end{equation}

Using $\int_{\eta}^{\infty}f_l^2(\lambda x)x^2dx=\frac{1}{2}\eta^3[f_{l-1}(\lambda \eta)f_{l+1}(\lambda \eta)-f_l^2(\lambda \eta)]$, where $\lambda$ and $\eta$ are real positive parameters, $\psi_ k(\vec{r})$ is now normalized in the whole space

\begin{equation}\label{renorma}
\int_{space}\psi_k^*(\vec{r})\psi_k(\vec{r})d^3r=1 \Rightarrow \int_{V_{\Xi}}\psi_k^*(\vec{r})\psi_k(\vec{r})d^3r\leq1.
\end{equation}

The charge of a SC $\Xi$ is equal to $Z_{\Xi}=Z-N_{\Xi,b}$. This enables us to define the real charge of a SC as the effective number of free electrons contained in the WS sphere

\begin{equation}
Z_{\Xi}^{real}=Z-N_{\Xi,b}(|\vec{r}|\leq r_{\Xi})\geq Z-N_{\Xi,b}(|\vec{r}|\leq +\infty)=Z_{\Xi}.
\end{equation}

\begin{figure}
\begin{center}
\includegraphics[width=8.6cm, trim=0 0 0 -60]{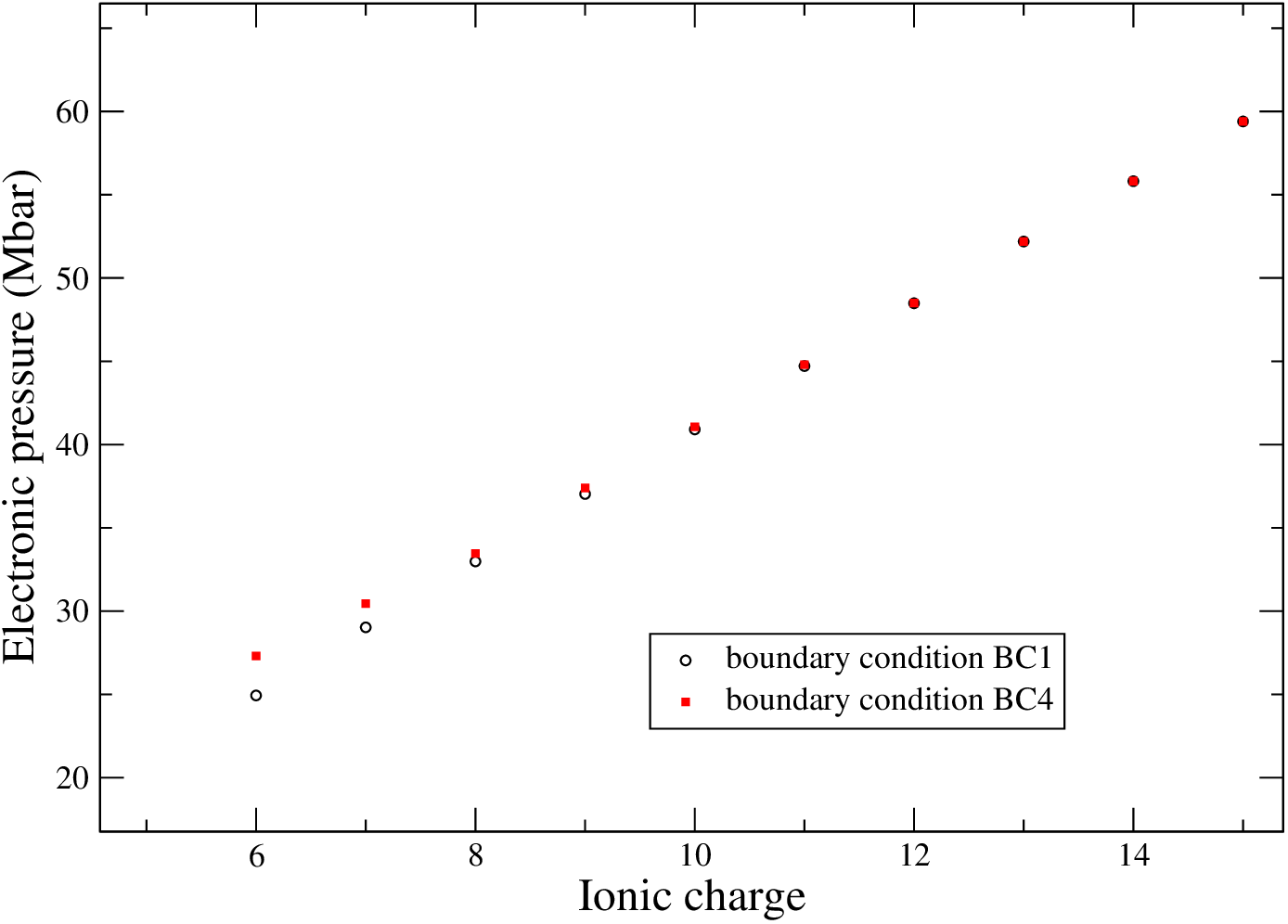}
\end{center}
\caption{\label{fig3}Pressure of ionic species for a vanadium plasma (V, Z=23) at T=100 eV and $\rho$=2.5 g/cm$^3$ with boundary conditions BC1 and BC4.}
\end{figure}

Calculations with this normalization in the whole space are referred as BC4. The effect of neighbouring ions is dominated by their spatial arrangement around a given ion. Thus, the effect of neighbouring ionic perturbers is different on each ion. It appears to be difficult to take into account the effect of surrounding ions on electronic structure. However, the description of the plasma within the ion-sphere model is a first step toward the treatment of ionic correlations, and the possibility of extension of a wavefunction outside the WS sphere can be a powerful tool to improve it. Indeed, only external states are really affected by neighbouring ions. However, a better insight on ionic correlations would be obtained with the introduction of ionic radial distribution functions $g_{i}(r)$. In the present case where ions are simply described by WS spheres, one has $g_{\Xi,i}(r)=\theta(r-r_{\Xi})$, $\theta$ being Heaviside function. Figure \ref{fig2} shows the effective ionic charges of a vanadium plasma at T=100 eV and $\rho$=2.5 g/cm$^3$. Only the small-size ions (lowest charge states) are affected. Figure \ref{fig3} shows pressure versus density for an aluminum plasma with boundary condition BC1 and BC4. Differences follow the tendency of figure \ref{fig2}.

\begin{figure}
\begin{center}
\includegraphics[width=8.6cm, trim=0 0 0 -60]{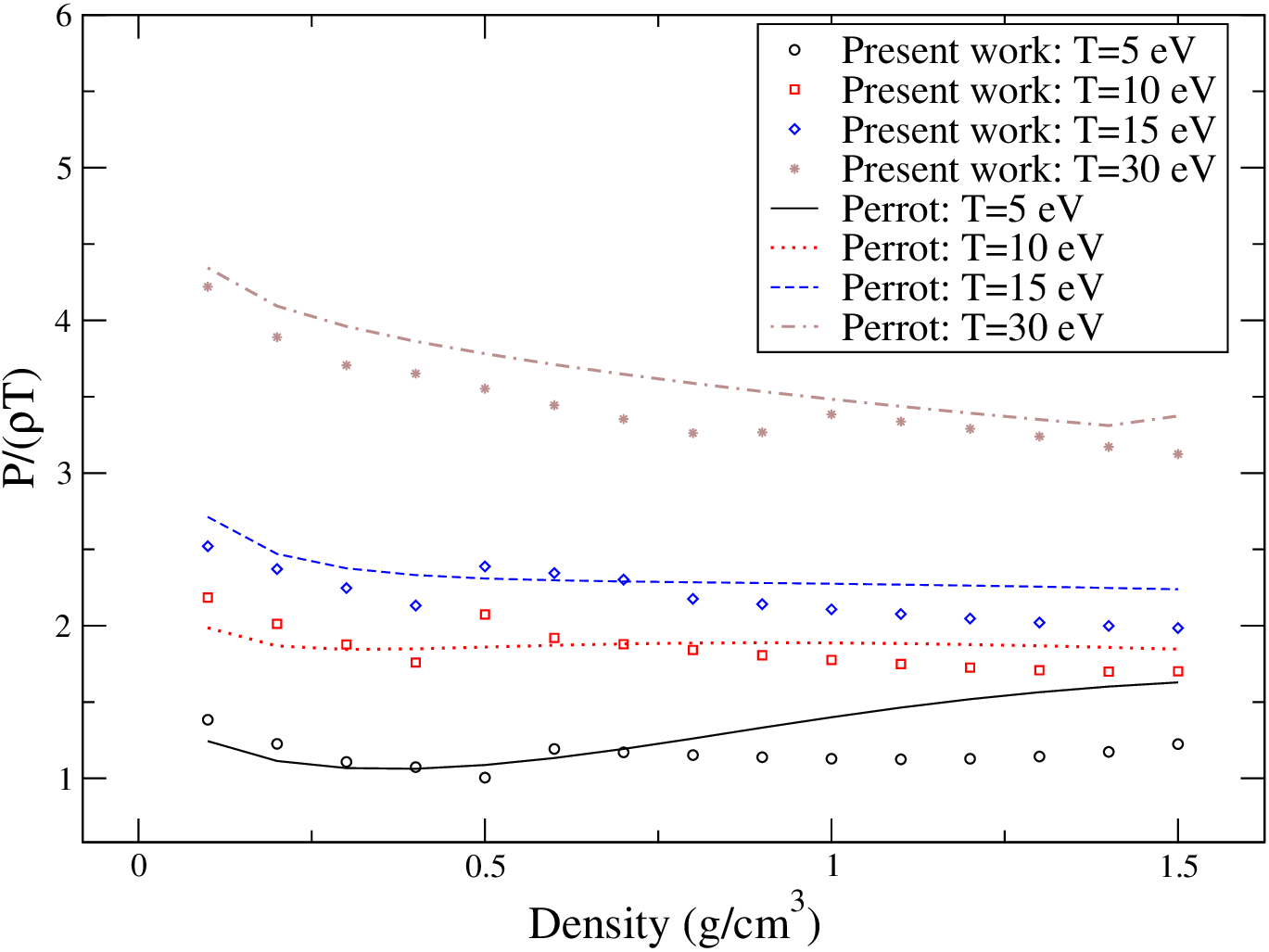}
\end{center}
\caption{\label{fig4}Pressure versus density for temperatures 5 eV, 10 eV, 15 eV and 30 eV for an aluminum plasma (Al, Z=13) calculated with our model and with Perrot's ``neutral pseudo-atom'' model.}
\end{figure}

\begin{figure}
\begin{center}
\includegraphics[width=8.6cm, trim=0 0 0 -60]{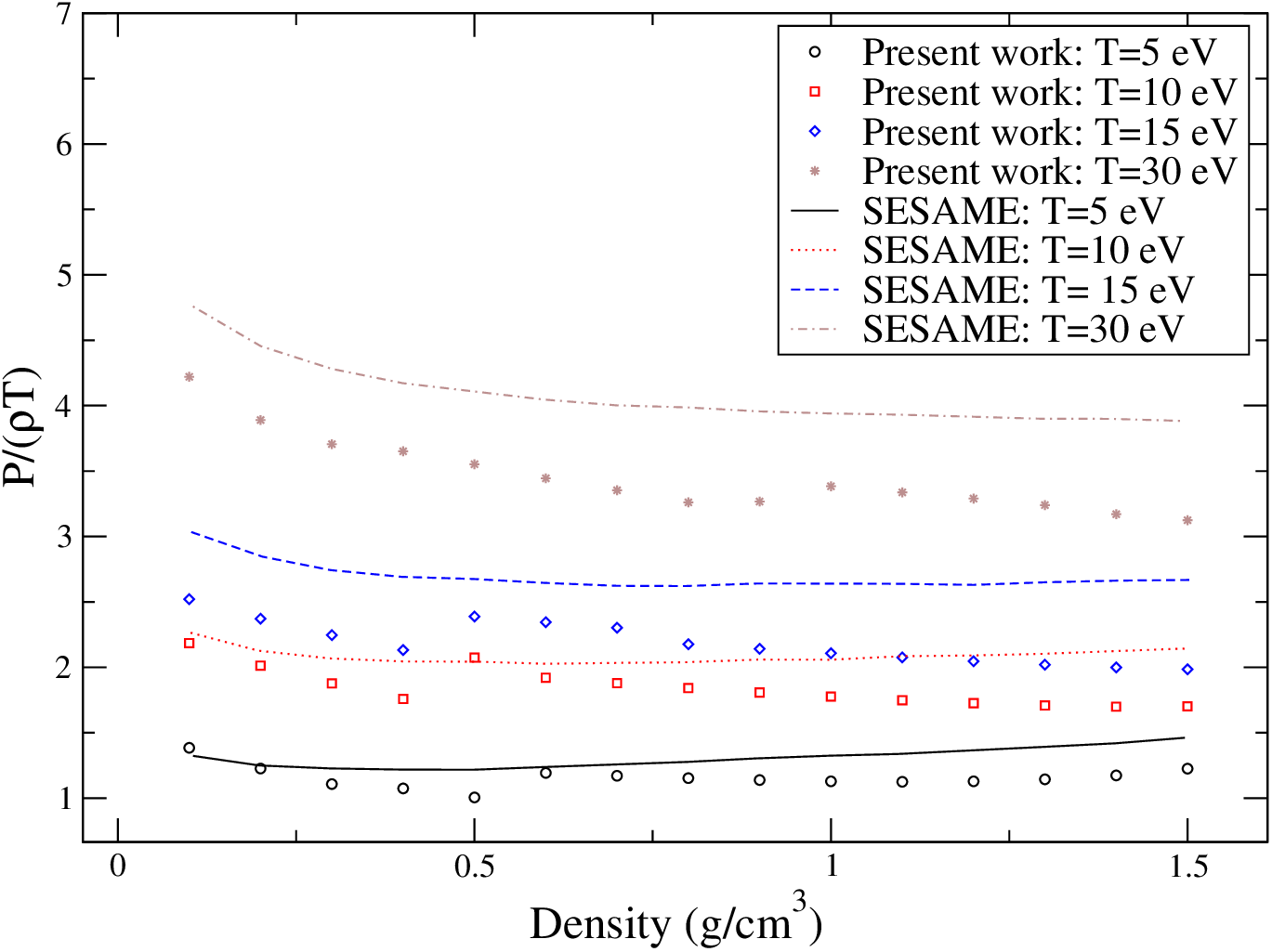}
\end{center}
\caption{\label{fig5}Pressure versus density for temperatures 5 eV, 10 eV, 15 eV and 30 eV for an aluminum plasma (Al, Z=13) calculated with our model and with ``Sesame'' EOS.}
\end{figure}

\begin{table}
\caption{\label{tab3}Pressures and densities for a vanadium plasma with ions from $V^{5+}$ to $V^{13+}$ at T=100 eV and $\rho$=1.7 g/cm$^3$.}
\begin{tabular}{|c|c|c|c|c|}\hline
Ion charge & 5 & 6 & 7 & 8\\\hline\hline
$P_{tot}$ (Mbar) & 15.17 & 17.83 & 20.58 & 23.37\\\hline
$\rho$ (g/cm$^3$) & 1.7 & 1.7 & 1.7 & 1.7\\\hline
\end{tabular}\\\\\\
\begin{tabular}{|c|c|c|c|c|}\hline
9 & 10 & 11 & 12 & 13\\\hline\hline
26.05 & 28.75 & 31.42 & 34.07 & 36.68\\\hline
1.7 & 1.7 & 1.7 & 1.7 & 1.7\\\hline
\end{tabular}
\end{table}

Checking the predictions of our model concerning the modeling of ions would require experiments with temperatures typically larger than 5 $eV$ and densities of the order of 1 g/cm$^3$. Unfortunately, even if a few experimental values do exist, the error bars are too large to see the difference between AA models and a SC model like ours. Moreover, we provide only the contribution of the electrons to the EOS, and building a complete EOS would require a model for the ions and a special treatment for low temperatures. 

However, it is instructive to compare our results with theoretical predictions from a model including ionic correlations. The Neutral Pseudo Atom model proposed by F. Perrot \cite{per2} combines an average-atom full self-consistent Kohn-Sham treatment of the electrons with a description of ions using the classical theory of liquids. For the electrons, the Kohn-Sham-Mermin equations are solved for a ``pseudo-atom'' embedded in a jellium of charges with a cavity. With the corresponding pseudo-potential, the ion-ion interaction is computed, in the Modified HyperNetted Chain (MHNC) approximation. The free energy consists of three parts:

\begin{equation}
F=Z^*f+\Delta F+F_{12},
\end{equation}

where $f$ is the free energy of an electron in a uniform interacting electron gas, $\Delta F$ is the free energy required to embed the atom in the jellium and $F_{12}$ is the free energy of ions. This treatment of ions constitutes the main difference with our approach, which relies on the re-construction of ions as groups of superconfigurations. In a sense, ionic correlations are not explicitly included in our model. Figure \ref{fig4} represents comparisons between excess pressures calculated from our approach and from Perrot's model for different densities between 0.1 g/cm$^3$ and 1.5 g/cm$^3$ and temperatures of 5 eV, 10 eV, 15 eV and 30 eV. More precisely, the quantity which is plotted is 

\begin{equation}
\frac{P_{ex}}{\rho T}=\frac{P}{\rho T}-1,
\end{equation}

where $P_{ex}$ is excess pressure and $P$ total pressure. Figure \ref{fig5} presents comparisons between our model and Sesame EOS in the same conditions as figure \ref{fig4}. The Sesame library, developed at Los Alamos Laboratory, is widely used for numerical simulations. In the region of interest, Sesame consists in an interpolation between the results of several theories valid in adjacent domains. At high density, Sesame relies on liquid metal perturbation theory with electronic excitations up to T=20 eV, and on TF theory with quantum and ionic corrections fot T $>$ 20 eV. For the low densities, Saha model is used for temperatures up to 20 eV, and an activity expansion (ACTEX) \cite{rog} based on static screened potentials is used for higher temperatures.

Concerning the effect of the shell structure on Hugoniot shock adiabats, at ultra-high pressure, laser-driven shock-wave experiments using high-power lasers (``National Ignition Facility'' in the USA or ``Laser M\'{e}gajoule'' in France) may provide experimental points in the future.

\section{\label{tow}Towards a variational formalism for EOS calculation}

\begin{figure}
\begin{center}
\includegraphics[width=8.6cm, trim=0 0 0 -60]{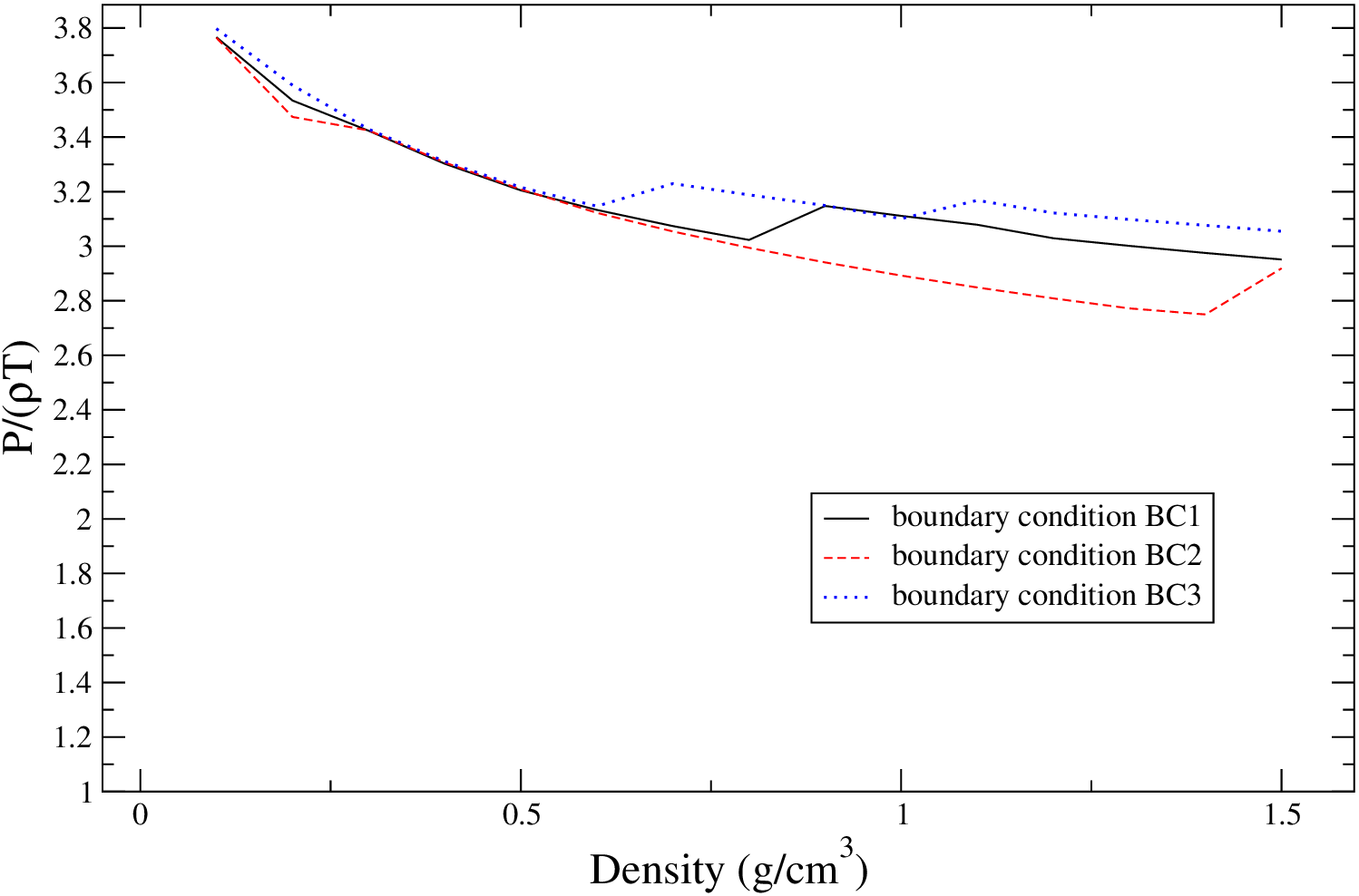}
\end{center}
\caption{\label{fig6}Pressure versus density for boundary conditions BC1, BC2 and BC3 for an aluminum plasma at T=25 eV.}
\end{figure}

The common definition of EOS consists in the study of pressure, internal energy and their variation with respect to temperature and volume (or matter density). In the present work, for a sake of simplicity, focus will only be put on pressure versus density. Table \ref{tab5} illustrates the fact that when density increases, a dissolution of successive orbitals into the continuum occurs (pressure ionization). Aluminum is interesting because it is often considered as a precise ``standard'' for EOS studies. Figure \ref{fig6} shows pressure versus density in the case of an aluminum plasma at T=25 eV for boundary conditions BC1, BC2 and BC3. For all boundary conditions, the curves exhibit a discontinuity, corresponding to pressure ionization of orbital 3p. For practical reasons, discontinuities are more obvious for function: $PV_{ws}/T\propto P/(\rho T)$, where $V_{ws}=4\pi r_{ws}^3/3$. Each ion in the plasma is surrounded by a unique spatial arrangement of the perturbing charges. Therefore, disappearance of a given energy level in an ensemble of ions is described by the fraction of ions that are perturbed strongly enough by their neighbours. The way an electron state is perturbed by the plasma environment is governed by continuous parameters, such as the distance to the neighbouring ionic species or spatial distributions of microfields. One possibility to correct such a non-physical behaviour would be to abandon the TF description of free electrons and to use a quantum treatment of free electrons, together with the inclusion of shape resonances \cite{mor,ble3,pai3}. A shape resonance is a peek in the free-electron density of states appearing simultaneously with the disappearing of a bound state into the continuum. The same approach has already been applied within the superconfiguration approximation \cite{pai3}. It provides continuous thermodynamic functions, but leads to a high numerical cost.

\subsection{Virial theorem}

Starting from Schr\"odinger equation, one can proove that the Virial theorem in the general case (\textit{i.e.} without any assumption about the boundary values of the wavefunctions) has the following form (see Appendix A):

\begin{equation}\label{vira}
K_1+K_2+E_{pot}=3PV,
\end{equation}

where $K_1=\frac{1}{2}\int_{V_{\Xi}}\vec{\nabla} \psi^*(\vec{r})\vec{\nabla}\psi(\vec{r})d^3r$ and $K_2=-\frac{1}{2}\int_{V_{\Xi}}\psi^*(\vec{r})\Delta\psi(\vec{r})d^3r$. It is particularly interesting to note that if all the one-electron wavefunctions obey Neumann boundary condition \cite{mor} $\vec{\nabla}\psi(\vec{r}_{\Xi})=0$ or Dirichlet boundary condition \cite{mor} $\psi(\vec{r}_{\Xi})=0$, then $K_1=K_2=E_{kin}$ and equation (\ref{vira}) takes the form \cite{mor,per}

\begin{equation}\label{virb}
2E_{kin}+E_{pot}=3PV 
\end{equation}

which is equivalent to the form of the Virial theorem for free electrons described within the TF approximation. Taking exchange-correlation effects into account, equation (\ref{vira}) reads \cite{leg}

\begin{equation}\label{virxc}
K_1+K_2-3V_{\Xi}n_{\Xi}(r_{\Xi})v_{xc}(r_{\Xi})+\int_0^{r_{\Xi}}v_{xc}(r)\frac{d(r^3n_{\Xi}(r))}{dr}4\pi dr=3PV.
\end{equation}

Appendix B shows that expression (\ref{vira}) of the Virial theorem leads to expression (\ref{expresso}) of electronic pressure. The use of Virial theorem (\ref{virb}) would lead to

\begin{eqnarray}\label{expresso2}
P_{\Xi,b}&=&\frac{1}{4\pi r_{ws}^2}\sum_{n,l}(2l+1)f_{nl}\Big\{\Big[\frac{dR_{nl}}{dr}(r_{\Xi})\Big]^2-\frac{1}{r_{\Xi}}R_{nl}(r_{\Xi})\frac{dR_{nl}}{dr}(r_{\Xi})\nonumber \\
&&-R_{nl}(r_{\Xi})\frac{d^2R_{nl}}{dr^2}(r_{\Xi})\Big\},
\end{eqnarray}

which can be written

\begin{eqnarray}
P_{\Xi,b}&=&\frac{1}{4\pi}\sum_{n,l}2(2l+1)f_{nl}[R_{nl}(r_{\Xi})]^2\times\nonumber \\
&&\Big[\frac{D_{nl}^2+D_{nl}-l(l+1)}{r_{\Xi}^2}+2(\epsilon_{nl}-v(r_{\Xi}))\Big],
\end{eqnarray}

which is different from expression (\ref{expdnl}): one finds $D_{nl}^2+D_{nl}-l(l+1)$ instead of $D_{nl}^2+2D_{nl}-l(l+1)$. Expresion (\ref{vira}) of Virial theorem and formula (\ref{expdnl}) for pressure hold for any boundary condition of the wavefunction.

\begin{figure}
\begin{center}
\includegraphics[width=8.6cm, trim=0 0 0 -60]{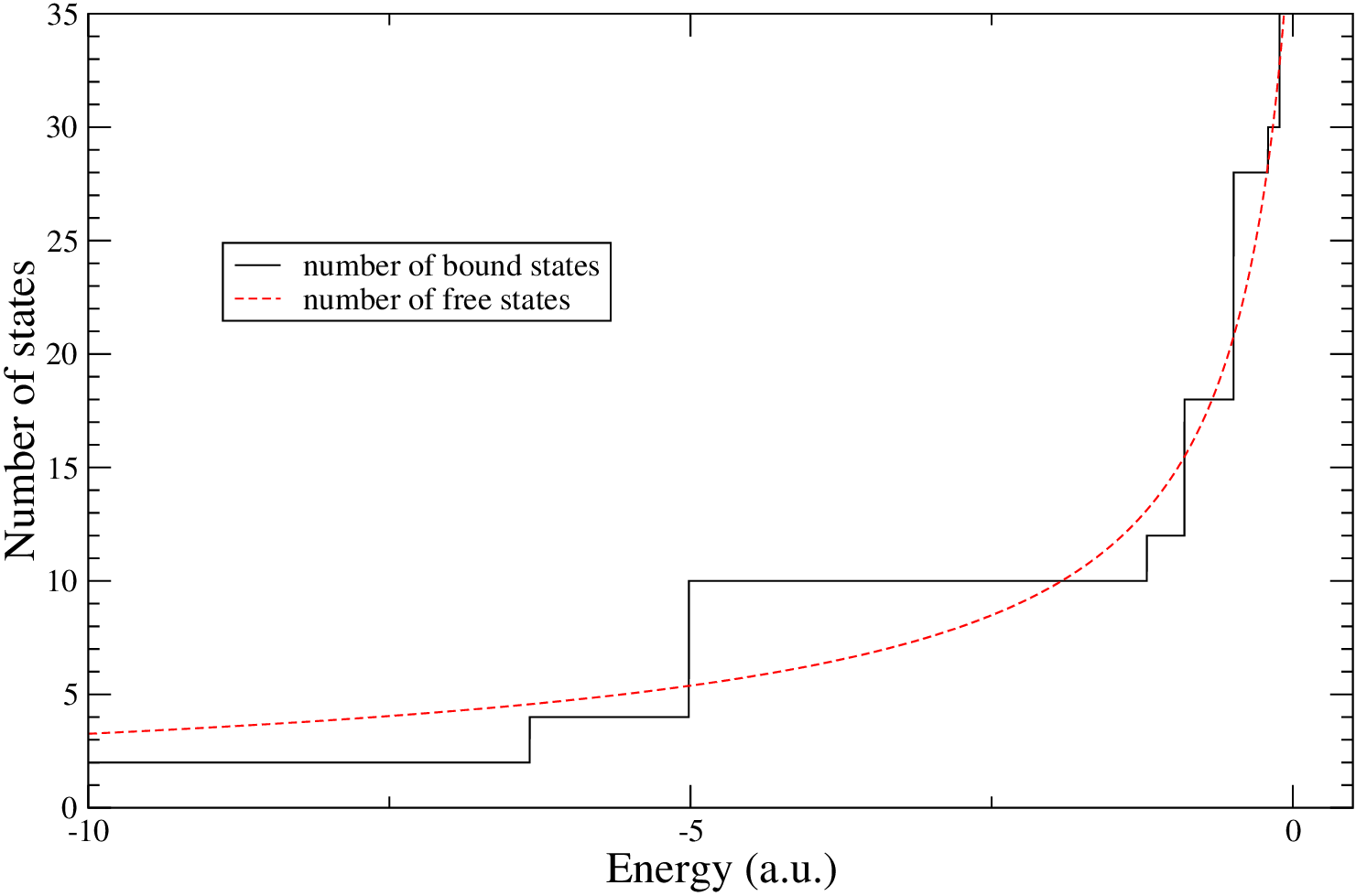}
\end{center}
\caption{\label{fig7}Number of states calculated in the TF approximation and quantum-mechanically for an aluminum plasma at T=25 eV and $\rho$=0.1 g/cm$^3$.}
\end{figure}

\begin{table}
\caption{\label{tab4}Pressures and densities for a vanadium plasma with ions from $V^{5+}$ to $V^{13+}$ at T=100 eV and $\rho$=1.7 g/cm$^3$ in the isobaric approach.}
\begin{tabular}{|c|c|c|c|c|}\hline
Ion charge & 5 & 6 & 7 & 8\\\hline\hline
$P_{tot}$ & 29.89 & 29.89 & 29.89 & 29.89\\\hline 
$\rho$ (g/cm$^3$) & 3.28 & 2.77 & 2.47 & 2.19\\\hline
\end{tabular}\\\\\\
\begin{tabular}{|c|c|c|c|c|}\hline
9 & 10 & 11 & 12 & 13\\\hline\hline
29.89 & 29.89 & 29.89 & 29.89 & 29.89\\\hline
1.96 & 1.77 & 1.61 & 1.48 & 1.37\\\hline
\end{tabular}
\end{table}

\subsection{Threshold of the continuum}\label{thr}

Close to energy zero, bound-electron band structure becomes larger, and continuum states acquire a more and more important resonant structure. As mentioned above, transition between both categories of electrons has to be continuous \cite{koh}. Sin'ko \cite{sin} and Nikiforov \emph{et al.} \cite{nik} have suggested, in the case of an AA calculation, to introduce a threshold for free-electron energy. In other words, free electrons are not defined as electrons with positive energy, but with energy greater than a non-zero value. Nikiforov \emph{et al.} \cite{nik} have shown, in the framework of a modified Hartree-Fock-Slater model, that the determination of the threshold of the continuum stems from a variational calculation. In the present work, this property is extended to a plasma described in the framework of density functional theory in the SC approximation. The number of bound electrons of a SC is an integer number, which leads to the determination of Lagrange multiplier $\mu_{\sigma}$. Inside the self-consistent calculation, all occupation factors are independently adjusted. Such an appoximation is equivalent to a translation of the origin of bound-electron energies. Thus, one can consider that the ``effective'' energies, \textit{i.e.} the energies that can be compared to the continuum threshold $\epsilon_t^{\Xi}$ are defined by: $\forall k \in \sigma, \tilde{\epsilon}_k=\epsilon_k-\mu_{\sigma}$. Bound states are defined by $\tilde{\epsilon}_k\leq \epsilon_t^{\Xi}$ and free states by $\epsilon\geq \epsilon_t^{\Xi}$. The grand potential of the entire plasma can be written in the following form

\begin{eqnarray}\label{grandpotcomp}
\Omega_{\Xi} &=&\sum_{\sigma\in\Xi}\sum_{k\in\sigma}g_kf_k\int_{V_{\Xi}}\psi_k^*(\vec{r})[-\frac{1}{2}\vec{\nabla}^2]\psi_k(\vec{r})\theta(\epsilon_t^{\Xi}-\tilde{\epsilon}_k)d^3r\nonumber \\
&&+2\int_{V_{\Xi}}\int_{p\geq p_0(r)}[1-\theta(\epsilon_t^{\Xi}-\epsilon)]n(r,\vec{p})\frac{p^2}{2}\frac{d^3rd^3p}{(2\pi)^3}\nonumber \\
&&+\int_{V_{\Xi}}n_{\Xi}(r)v_e(r)d^3r+\frac{1}{2}\int_{V_{\Xi}}\int_{V_{\Xi}}\frac{n_{\Xi}(r)n_{\Xi}(r')}{|\vec{r}-\vec{r}'|}d^3rd^3r'\nonumber \\ 
&&-\mu_{\Xi}[\int_{V_{\Xi}}n_{\Xi}(r)d^3r-Z]\nonumber \\ 
&&+T\sum_{\sigma\in\Xi}\sum_{k\in\sigma}\theta(\epsilon_t^{\Xi}-\tilde{\epsilon}_k)g_k[f_k\ln f_k+(1-f_k)\ln (1-f_k)] \nonumber \\
&&+2T\int_{V_{\Xi}}\int_{p\geq p_0(r)}[n(r,\vec{p})\ln n(r,\vec{p})\nonumber \\
&&+(1-n(r,\vec{p}))\ln (1- n(r,\vec{p}))][1-\theta(\epsilon_t^{\Xi}-\epsilon)]\frac{d^3rd^3p}{(2\pi)^3}\nonumber \\
&&+\int_{V_{\Xi}}f_{\Xi,xc}[n_{\Xi}]d^3r,
\end{eqnarray}

where $v_e$ is the external potential, including the potential of the nucleus. The total electron density is given by

\begin{eqnarray}
n_{\Xi}(r)&=&\sum_{\sigma\in\Xi}\sum_{k\in\sigma}g_kf_k\psi_k^*(\vec{r})\psi_k(\vec{r})\theta(\epsilon_t^{\Xi}-\tilde{\epsilon}_k)\nonumber \\
&&+2\int_{p\geq p_0(r)}\frac{p^2}{2}n(r,\vec{p})[1-\theta(\epsilon_t^{\Xi}-\epsilon)]\frac{d^3rd^3p}{(2\pi)^3}
\end{eqnarray}

and $p_0$ is a fonction of $r$ defined by $p_0^2(r)=\max[0,-2v(r)]$. Infinitesimal variation of equation (\ref{granpot}), \textit{i.e.} $\delta\tilde{\Omega}=0$, has to be carried out using expression (\ref{grandpotcomp}) of $\Omega_{\Xi}$. Since one has

\begin{equation}\label{schro}
[-\frac{1}{2}\Delta+v(r)]\psi_k=\epsilon_k\psi_k, 
\end{equation}

variation with respect to $\delta \psi_k^*$ gives 

\begin{equation}\label{energy}
\epsilon_k=\mu_{\sigma}+\mu_{\Xi}-\frac{\lambda_k}{g_kf_k}, 
\end{equation}

variation with respect to $\delta f_k$ gives 

\begin{equation}\label{occupation}
f_k=\frac{1}{1+\exp[\frac{\tilde{\epsilon}_k-\mu_{\Xi}}{T}]}=\frac{1}{1+\exp[\frac{\epsilon_k-\mu_{\sigma}-\mu_{\Xi}}{T}]}, 
\end{equation}

variation with respect to $\delta n$ gives 

\begin{equation}\label{density}
n(r,\vec{p})=\frac{1}{1+\exp[\frac{\frac{p^2}{2}+v(r)-\mu_{\Xi}}{T}]},
\end{equation}

variation with respect to $V_{\Xi}$ gives

\begin{equation}
P_{\Xi}\doteq-\frac{\partial\tilde{\Omega}_{\Xi}}{\partial V_{\Xi}}\Bigl|_{T}=P,
\end{equation}

variation with respect to $V_{\Xi}$ gives

\begin{equation}
W_{\Xi}\propto\exp [-\frac{\tilde{\Omega}_{\Xi}+PV_{\Xi}}{T}],
\end{equation}

and variation with respect to function $\theta$ gives

\begin{eqnarray}\label{condi}
\sum_{\sigma\in\Xi}\sum_{k\in\sigma}g_k\delta \theta(\epsilon_t^{\Xi}-\tilde{\epsilon}_k)[\tilde{\epsilon}_k-\mu_{\Xi}+T\ln f_k]\nonumber \\
-2\int_{V_{\Xi}}\int_{p\geq p_0(r)}\delta\theta(\epsilon_t^{\Xi}-\epsilon)[\epsilon-\mu_{\Xi}+T\ln n(r,\vec{p})]\frac{d^3p}{(2\pi)^3}d^3r=0.
\end{eqnarray}

Fonctions $\delta \theta(\epsilon_t^{\Xi}-\tilde{\epsilon}_k)$ and $\delta\theta(\epsilon_t^{\Xi}-\epsilon)$ are non-zero if and only if $\tilde{\epsilon}_k=\epsilon_k-\mu_{\sigma}=\epsilon_t^{\Xi}$ and $\epsilon=p^2/2+v(r)=\epsilon_t^{\Xi}$. Equation (\ref{condi}) can then be re-written 

\begin{equation}\label{condi2}
\Upsilon(\epsilon_t^{\Xi})\delta[\sum_{\sigma\in\Xi}\sum_{k\in\sigma}g_k\theta(\epsilon_t^{\Xi}-\tilde{\epsilon}_k)-2\int_{V_{\Xi}}\int_{p\geq p_0(r)}\theta(\epsilon_t^{\Xi}-\epsilon)\frac{d^3p}{(2\pi)^3}d^3r]=0, 
\end{equation}

where $\Upsilon(\epsilon_t^{\Xi})$ is defined as 

\begin{equation}\label{m}
\Upsilon(\epsilon_t^{\Xi})=\epsilon_t^{\Xi}-\mu_{\Xi}-T\ln [1+\exp[\frac{\epsilon_t^{\Xi}-\mu_{\Xi}}{T}]].
\end{equation}

If $\epsilon_t^{\Xi}-\mu_{\Xi}\gg T$, \textit{i.e.} if the electron-gas is totally degenerate, then $\Upsilon(\epsilon_t^{\Xi})\approx 0$ and condition (\ref{condi2}) is fulfilled. In more usual situations, condition (\ref{condi2}) becomes

\begin{equation}\label{condi3}
\sum_{\sigma\in\Xi}\sum_{k\in\sigma}(2l_k+1)\theta(\epsilon_t^{\Xi}-\tilde{\epsilon}_k)=\int_{V_{\Xi}}\int_{p\geq p_{0,t}^{\Xi}(r)}\frac{d^3p}{(2\pi)^3}d^3r.
\end{equation}

The free-electron momentum obeys the inequality $p^2/2\geq\epsilon_t^{\Xi}-v(r)$. The threshold momentum $p_{0,t}^{\Xi}(r)$ is therefore defined by $[p_{0,t}^{\Xi}(r)]^2=\max[0,2(\epsilon_t^{\Xi}-v(r))]$. 

\subsection{Determination of thermodynamic pressure as a derivative of grand potential}

The total pressure of SC $\Xi$ is given by $P_{\Xi}=-\frac{\partial \Omega_{\Xi}}{\partial V_{\Xi}}\Bigl|_{T}=-\frac{1}{4\pi r_{\Xi}^2}\frac{\partial \Omega_{\Xi}}{\partial r_{\Xi}}\Bigl|_{T}$. Let us calculate infinitesimal variation of equation (\ref{grandpotcomp}) with respect to $r_{\Xi}$. It is important to mention that in the expression of grand potential, wavefunctions $\psi_k$ and Heavyside function $\theta$ depend on $r_{\Xi}$ in an implicit way. The total pressure resulting from derivation of grand potential (\ref{grandpotcomp}) with respect to volume \cite{sin} is given by the formula 

\begin{eqnarray}\label{exppress}
P_{\Xi}&=&P_{\Xi,b}+P_{\Xi,f}+P_{\Xi,xc}\nonumber \\
&=&\frac{1}{4\pi^2}\sum_{n,l}\theta(\epsilon_t^{\Xi}-\tilde{\epsilon}_{nl})(2l+1)f_{nl}\Big[\Big(\frac{dR_{nl}}{dr}(r_{\Xi})\Big)^2-R_{nl}(r_{\Xi})\frac{d^2R_{nl}}{dr^2}(r_{\Xi})\Big]\nonumber \\
&&+\frac{2\sqrt{2}T^{5/2}}{3\pi^2}\int_{x_{0,t}^{\Xi}(r)}^{+\infty}\frac{x^{3/2}}{1+\exp[x-\frac{v(r)-\mu_{\Xi}}{T}]}dx+(n_{\Xi}\frac{\partial f_{\Xi,xc}}{\partial n_{\Xi}}-f_{\Xi,xc})_{r_{\Xi}}\nonumber \\
&&+\frac{T}{4\pi r_{\Xi}^2}\Big[\frac{\epsilon_t^{\Xi}-\mu_{\Xi}}{T}-\ln[1+\exp[\frac{\epsilon_t^{\Xi}-\mu_{\Xi}}{T}]]\Big]\times\nonumber \\
&&\frac{\partial}{\partial r_{\Xi}}
[\frac{8\sqrt{2}}{3\pi}\int_{0}^{r_{\Xi}}[y_{0,t}^{\Xi}(r)]^{3/2}r^2dr-
\sum_{n,l}\theta(\epsilon_t^{\Xi}-\tilde{\epsilon}_{nl})2(2l+1)], 
\end{eqnarray}

where $x_{0,t}^{\Xi}(r)$ is equal to $y_{0,t}^{\Xi}(r)/T$. According to condition (\ref{condi3}), the last term is zero, and equation (\ref{exppress}) is equivalent to equation (\ref{expresso}).

\subsection{Non-unicity of the choice of the continuum threshold}

In a first approximation and for a sake of simplicity, it is interesting to consider the AA case. Then, there is only one threshold energy for the plasma and all populations of bound levels are fractional \cite{roz}. Equation (\ref{condi3}) has more than one solution. Figure \ref{fig7} shows the density of states calculated in the TF approximation and in the framework of quantum mechanics and illustrates the non-unicity of the continuum threshold. One solution consists, following Nikiforov \emph{et al.} \cite{nik}, in numbering all the roots (in an increasing order), and fixing, for a given element, the number of the choosen root. The most difficult part of such procedure stems from the definition of a criterium, stating the number of the root. It appears that the only way to ensure continuity implies not only to choose the same root-number for all densities, but also a constant number of bound levels. Unfortunately, this is not acceptable, since if the level which is supposed to disappear in the considered density range is rejected into the continuum from the beginning, the continuity of pressure is obvious. 

In the SC approximation, each SC has its own continuum threshold $\epsilon_t^{\Xi}$, which gives an additional degree of freedom, even if the determination of the root is still problematic. In the isobaric approach, one can wonder whether the fact that each ion has its own continuum threshold is in contradiction or not with the fact that all ions should have the same electronic environment.

\begin{table}
\caption{\label{tab5}Coupling parameter and binding energies of orbitals for a vanadium plasma at T=100 eV and different values of density. The binding energy (absolute value of the energy of the last orbital) is expressed in atomic units (a.u.) defined by $m=\hbar=e=1$.}
\begin{tabular}{|c|c|c|c|c|c|c|c|}\hline
Density (g/cm$^3$) & 0.1 & 0.2 & 0.3 & 0.4 & 0.5 & 0.6 & 0.7\\\hline\hline
$\Gamma_{ii}$ & 2.44 & 2.77 & 3.00 & 3.17 & 3.32 & 3.44 & 3.55\\\hline
$E_{binding}$(a.u.) & 0.109 & 0.049 & 5.87.10$^{-4}$ & 0.362 & 0.276 & 0.208 & 0.153\\\hline
Last orbital & 4p & 4s & 4s & 3p & 3p & 3p & 3p\\\hline
\end{tabular}\\\\\\
\begin{tabular}{|c|c|c|c|c|c|c|c|}\hline
0.8 & 0.9 & 1.0 & 1.1 & 1.2 & 1.3 & 1.4 & 1.5\\\hline\hline
3.64 & 3.72 & 3.80 & 3.86 & 3.87 & 3.92 & 3.72 & 3.99\\\hline
0.107 & 0.070 & 0.038 & 0.013 & 0.216 & 0.187 & 0.161 & 0.137\\\hline
3p & 3p & 3p & 3p & 3s & 3s & 3s & 3s\\\hline
\end{tabular}
\end{table}

\clearpage

\section{Conclusion and perspectives}

A self-consistent model for the calculation of atomic shell structure and pressure in hot dense plasmas was presented. In this approach, ions are described by superconfigurations, in order to explore atomic physics beyond the AA model without having to take the tremendous number of electronic configurations into account. The model can be extended to an isobaric context, in which all ions have the same electronic environment (same boundary pressure) and different volumes. The possibility of extension of bound-electron wavefunctions outside the WS sphere is a first attempt to consider ionic correlations in an undirect and qualitative way, and a useful indicator of the real charges of ions. In most of quantum self-consistent-field models, either in the AA or SC cases, bound electrons are described quantum-mechanically and free electrons in the TF approximation, in order to avoid a quantum treatment of continuum states which leads to a high computational time.

The uncertainty concerning the boundary values of the bound-electron wavefunctions leads to differences in the thermodynamic quantities. It was shown that the Virial theorem for bound electrons without any assumption concerning the boundary conditions is different from the one of a free-electron gas. This has an impact on the stress-tensor pressure formula determined from that specific Virial theorem. The pressure calculated from that generalized Virial theorem is equal to the pressure obtained as a derivative of the free energy.

In the hybrid model (quantum bound electrons and Thomas-Fermi free electrons), discontinuities appear in the thermodynamic functions when pressure ionization occurs. A formal condition was presented which should make the hybrid model variational, and therefore provide a suitable treatment of pressure ionization. It relies on the definition of a ``continuum threshold''for each superconfiguration.

The superconfiguration method relies on the calculation of partition functions (as explained in section \ref{sel}) in which interaction terms are averaged over all the configurations belonging to the superconfiguration. It is absolutely necessary to get an estimation of the error induced by such an \textit{ansatz}, and to try to incorporate, even in an approximate way, this contribution in the calculation of ion partition functions and therefore of all thermodynamic quantities.

\clearpage

\section{Appendix A: Virial theorem from Schr\"odinger equation}

One-electron wavefunctions in the potential $v$ satisfy Schr\"odinger equation
 
\begin{equation}
-\frac{1}{2}\vec{\nabla}^2\psi(\vec{r},t)+v(r)\psi(\vec{r},t)=i\frac{\partial
\psi(\vec{r},t)}{\partial t}\label{kkk}.
\end{equation}

The derivative of quantity

\begin{equation}
\Lambda_{\alpha}=\psi^*\hat{p_{\alpha}}\psi
\end{equation}

with respect to time $t$ gives

\begin{equation}
\frac{d\Lambda_{\alpha}}{dt}=\Big[\frac{\partial\psi^*}{\partial t}\hat{p_{\alpha}}\psi+
\psi^*\hat{p_{\alpha}}\frac{\partial\psi}{\partial t}].
\end{equation}

Equation (\ref{kkk}) enables one to write

\begin{equation}
\frac{d\Lambda_{\alpha}}{dt}=\frac{1}{2i}\{[\vec{\nabla}^2\psi^*\hat{p_{\alpha}}\psi-
\psi^*\hat{p_{\alpha}}\vec{\nabla}^2\psi]+i[\psi^*v\hat{p_{\alpha}}\psi-
\psi^*\hat{p_{\alpha}}v\psi]\}.
\end{equation}

Furthermore,

\begin{equation}
\frac{d\Lambda_{\alpha}}{dt}=\frac{1}{2i}\vec{\nabla}[\vec{\nabla}\psi^*\hat{p_{\alpha}}\psi-
\psi^*\hat{p_{\alpha}}\vec{\nabla}\psi]-i\psi^*[\hat{p_{\alpha}},v]\psi.
\end{equation}

Using the relation $\hat{p_{\alpha}}=\frac{1}{i}\frac{\partial}{\partial x_{\alpha}}$ enables one to write

\begin{equation}\label{zozo}
\frac{d\Lambda_{\alpha}}{dt}=\frac{1}{2i}\sum_j\frac{\partial}{\partial x_{\alpha}}\Big[\frac{\partial\psi^*}{\partial
x_j}\frac{\partial\psi}{\partial x_{\alpha}}-\psi^*\frac{\partial^2\psi}{\partial x_{\alpha}\partial x_j}\Big]-\psi^*\frac{\partial v}{\partial x_{\alpha}}\psi.
\end{equation}

In a stationary regime ($\frac{d\Lambda_{\alpha}}{dt}=0$), equation (\ref{zozo}) becomes

\begin{equation}\label{koala}
\frac{1}{2i}\sum_j\frac{\partial}{\partial x_{\alpha}}\Big[\frac{\partial\psi^*}{\partial
x_j}\frac{\partial\psi}{\partial x_{\alpha}}-\psi^*\frac{\partial^2\psi}{\partial x_{\alpha}\partial x_j}\Big]=\psi^*\frac{\partial v}{\partial x_{\alpha}}\psi.
\end{equation}

In order to take into account FD statistics, relation (\ref{koala}) becomes

\begin{equation}\label{eqx}
\frac{1}{2i}\sum_{k,j}f_k\frac{\partial}{\partial x_{\alpha}}\Big[\frac{\partial\psi_k^*}{\partial
x_j}\frac{\partial\psi_k}{\partial x_{\alpha}}-\psi_k^*\frac{\partial^2\psi_k}{\partial x_{\alpha}\partial x_j}\Big]=\sum_kf_k\psi_k^*\frac{\partial v}{\partial x_{\alpha}}\psi_k,
\end{equation}

where $k$ represents an electronic state and $f_k=\frac{1}{\exp[(\epsilon_k-\mu)/T]+1}$. It is easy to recognize the tensor

\begin{equation}
P_{ji}=\frac{1}{2}\sum_kf_k\Big[\frac{\partial\psi_k^*}{\partial x_j}
\frac{\partial\psi_k}{\partial x_{\alpha}}-\psi_k^*\frac{\partial^2\psi_k}{\partial x_{\alpha}\partial x_j}\Big],
\end{equation}

which enables one to rewrite equation (\ref{eqx}) as

\begin{equation}\label{prec}
\sum_j\frac{\partial}{\partial x_{\alpha}}P_{j\alpha}=-\sum_kf_k\psi_k^*\frac{\partial
v}{\partial x_{\alpha}}\psi_k.
\end{equation}

Multiplying equation (\ref{prec}) by $x_{\alpha}$, summing over $\alpha$ and integrating over the volume leads to

\begin{equation}\label{zaza}
\sum_{\alpha,j}\int_{V}x_{\alpha}\frac{\partial P_{j\alpha}}{\partial x_j}d^3r=-\sum_kf_k\sum_{\alpha}\int_{V}x_{\alpha}\psi_k^*
\frac{\partial v}{\partial x_{\alpha}}\psi_k d^3r.
\end{equation}

Electronic density reads $n(\vec{r})=\sum_kf_k\psi_k\psi_k^*$. It is easy to recognize in equation (\ref{zaza}) the $\alpha^{th}$ coordinate of the electric field $\vec{E}$: $E_{\alpha}=-\frac{\partial v}{\partial x_{\alpha}}$. Equation (\ref{zaza}) thus reads

\begin{equation}
\sum_{\alpha,j}\int_{V}x_{\alpha}\frac{\partial P_{j\alpha}}{\partial
x_j}d^3r=\sum_{\alpha}\int_{V}x_{\alpha}E_{\alpha}\rho(\vec{r})d^3r
=\int_{V}\vec{r}.\vec{E}(\vec{r})n(\vec{r})d^3r=E_{pot},
\end{equation}

$E_{pot}$ being total potential energy. Furthermore, one has

\begin{equation}
\sum_{\alpha,j}\int_{V}x_{\alpha}\frac{\partial P_{j\alpha}}{\partial
x_j}d^3r=\sum_{\alpha,j}\int_{V}\frac{\partial}{\partial x_j}[x_{\alpha}P_{j\alpha}]d^3r
-\sum_{\alpha,j}\int_{V}P_{j\alpha}\frac{\partial x_{\alpha}}{\partial x_j}d^3r,
\end{equation}

which can be re-written

\begin{equation}
\sum_{\alpha,j}\int_{V}x_{\alpha}\frac{\partial P_{j\alpha}}{\partial
x_j}d^3r=\sum_{\alpha,j}\int_{\partial V}x_{\alpha}P_{j\alpha}dS_j
-\sum_{\alpha,j}\int_{V}P_{j\alpha}\delta_{\alpha,j}d^3r,
\end{equation}

$\partial V$ being the surface delimited by volume $V$. In the present case, $P_{\alpha j}=P\delta_{\alpha,j}$, which leads to 

\begin{equation}
\sum_{\alpha,j}\int_{V}x_{\alpha}\frac{\partial P_{j\alpha}}{\partial x_j}d^3r=P\sum_{\alpha,j}\int_{\partial V}x_{\alpha}\delta_{\alpha,j}dS_j
-\sum_{\alpha}\int_{V}P_{\alpha\alpha}d^3r.
\end{equation}

Introduction of quantity $P$ corresponds to a possible definition of pressure. At this stage, nothing indicates that this quantity is equal to the thermodynamic pressure obtained as a derivative of total grand potential. One has

\begin{equation}\label{bil}
P\sum_{\alpha,j}\int_{\partial V}x_{\alpha}\delta_{\alpha,j}dS_j
=P\int_{\partial V}\vec{r}.\vec{dS}=P\int_{V}(\vec{\nabla}.\vec{r})\vec{dr}
=3P\int_{V}\vec{dr}=3PV.
\end{equation}

The last term in equation (\ref{bil}) can be written

\begin{equation}
\sum_{\alpha}\int_{V}P_{\alpha\alpha}d^3r
=\sum_{\alpha}\frac{1}{2}\sum_kf_k\int_{V}\Big[\frac{\partial\psi_k^*}{\partial x_{\alpha}}
\frac{\partial\psi_k}{\partial x_{\alpha}}-\psi_k^*\frac{\partial^2\psi_k}{\partial x_{\alpha^2}}\Big]d^3r,
\end{equation}

or

\begin{equation}
\sum_{i}\int_{V}P_{\alpha\alpha}d^3r
=\frac{1}{2}\sum_kf_k\int_{V}|\vec{\nabla}\psi_k|^2d^3r
-\frac{1}{2}\sum_kf_k\int_{V}\psi_k^*\vec{\nabla}^2\psi_k d^3r.
\end{equation}

Finally one has $3PV=K_1+K_2+E_{pot}$ with

\begin{equation}
K_1=\frac{1}{2}\sum_kf_k\int_{V}|\vec{\nabla}\psi_k|^2d^3r \;\; ; \;\; K_2=-\frac{1}{2}\sum_kf_k\int_{V}\psi_k^*\vec{\nabla}^2\psi_k d^3r,
\end{equation}

which is the Virial theorem without any assumption concerning the boundary values of the wavefunctions.

\clearpage

\section{Appendix B: Calculation of bound-electron contribution to pressure from the Virial theorem}

The Virial theorem (see Appendix A) can be written $3PV=K_1+K_2+E_{pot}$, with

\begin{equation}
K_1=\frac{1}{2}\sum_kf_k\int_{V}|\vec{\nabla}\psi_k|^2d^3r \;\; ; \;\; K_2=-\frac{1}{2}\sum_kf_k\int_{V}\psi_k^*\vec{\nabla}^2\psi_k d^3r,
\end{equation}

where $\psi_k(r,\theta,\phi)=R_{nl}(r)Y_{lm}(\theta,\phi)\chi_s$. One has

\begin{eqnarray}
|\vec{\nabla}\psi(r,\theta,\phi)|^2&=&\Big[\frac{dR_{nl}(r)}{dr}\Big]^2|Y_{lm}(\theta,\phi)|^2+\frac{R_{nl}^2(r)}{r^2}\Big|
\frac{\partial Y_{lm}(\theta,\phi)}{\partial \theta}\Big|^2\nonumber\\
&&+\frac{R_{nl}^2(r)}{r^2\sin^2\theta}\Big|\frac{\partial Y_{lm}(\theta,\phi)}{\partial \phi}\Big|^2,
\end{eqnarray}

which, taking the spin degeneracy into account, leads to

\begin{eqnarray}
K_1&=&\sum_{n,l}f_{nl}\int_0^{\vec{r}_{ws}}\Big\{\Big[\frac{dR_{nl}(r)}{dr}\Big]^2\sum_{m=-l}^{l}|Y_{lm}(\theta,\phi)|^2+\frac{R_{nl}^2(r)}{r^2}\sum_{m=-l}^{l}
\Big|\frac{\partial Y_{lm}(\theta,\phi)}{\partial \theta}\Big|^2\nonumber \\
&&+\frac{R_{nl}^2(r)}{r^2\sin^2\theta}\sum_{m=-l}^{l}\Bigl|\frac{\partial Y_{lm}(\theta,\phi)}{\partial \phi}\Bigl|^2\Big\}\sin\theta d\theta d\phi dr
\end{eqnarray}

and

\begin{eqnarray}
K_2&=&-\sum_{n,l,m}f_{nl}\int_0^{\vec{r}_{ws}} |Y_{lm}(\theta,\phi)|^2[rR_{nl}(r)]\Big[\frac{d^2}{dr^2}[rR_{nl}(r)]\nonumber \\
&&-\frac{l(l+1)}{r^2}[rR_{nl}(r)]\Big]\sin\theta d\theta d\phi dr.
\end{eqnarray}

Using the relations

\begin{equation}\label{one}
\sum_{m=-l}^{l}|Y_{lm}(\theta,\phi)|^2=\frac{2l+1}{4\pi},
\end{equation}

\begin{equation}
\sum_{m=-l}^{l}\Big|\frac{\partial Y_{lm}(\theta,\phi)}{\partial
\theta}\Big|^2=\frac{(2l+1)}{4\pi}\frac{l(l+1)}{2}
\end{equation}

and

\begin{equation}
\sum_{m=-l}^{l}\Big|\frac{\partial Y_{lm}(\theta,\phi)}{\partial
\phi}\Big|^2=\frac{(2l+1)}{4\pi}\frac{l(l+1)}{2}\sin^2\theta,
\end{equation}

$K_1$ can be written

\begin{equation}
K_1=\sum_{n,l}(2l+1)f_{nl}\int_0^{r_{ws}}\Big\{\Big[\frac{dR_{nl}(r)}{dr}\Big]^2+\frac{l(l+1)}{2r^2}R_{nl}^2(r)\Big\}r^2dr
\end{equation}

and

\begin{eqnarray}
K_2&=&-\sum_{n,l}(2l+1)f_{nl}\int_0^{r_{ws}}[rR_{nl}(r)]\times\nonumber \\
&&\Big[\frac{d^2}{dr^2}[rR_{nl}(r)]-\frac{l(l+1)}{r^2}[rR_{nl}(r)]\Big]r^2dr.
\end{eqnarray}

Following the calculation of Legrand \emph{et al.} \cite{leg}, let us introduce auxiliary integral (the factor 2 is due to the spin degeneracy)

\begin{equation}\label{muse}
\tilde{L}_{nlm}=2f_{nl}\int_0^{\vec{r}_{ws}}r^3R_{nl}^2(r)|Y_{lm}(\theta,\phi)|^2\frac{dv}{dr}dr \sin\theta d\theta d\phi.
\end{equation}

Relation (\ref{one}) enables one to write expression $\sum_{n,l,m}\tilde{L}_{nlm}$ in the following form

\begin{eqnarray}
\sum_{n,l,m}\tilde{L}_{nlm}&=&K_2+\sum_{n,l}(2l+1)f_{nl}\frac{l(l+1)}{2}\int_0^{r_{ws}}\Big[\frac{R_{nl}(r)}{r}\Big]^2r^2dr\nonumber \\
&&-\sum_{n,l}(2l+1)f_{nl}r_{ws}^3R_{nl}^2(r_{ws})\nonumber\times\nonumber \\
&&\Big[\frac{D_{nl}^2+2D_{nl}-l(l+1)}{r_{ws}^2}+2[\epsilon_{nl}-v(r_{ws})]\Big]\nonumber \\
&&+\sum_{n,l}(2l+1)f_{nl}\int_0^{r_{ws}}\Big[\frac{dR_{nl}(r)}{dr}\Big]^2r^2dr.
\end{eqnarray}

Taking into account the fact that $\sum_{n,l,m}\tilde{L}_{nlm}=-E_{pot}$, $E_{pot}$ being potential energy, it is possible to write

\begin{eqnarray}
&&K_2+K_1-\sum_{n,l}(2l+1)r_{ws}^3R_{nl}^2(r_{ws})|Y_{lm}(\theta_{r_{ws}},\phi_{r_{ws}})|^2\times\nonumber \\
&&\Big[\frac{D_{nl}^2+2D_{nl}-l(l+1)}{r_{ws}^2}+2(\epsilon_{nl}-v(r_{ws}))\Big]=-E_{pot}
\end{eqnarray}

The Virial theorem being $E_{pot}+K_1+K_2=4\pi r_{ws}^3P$, one obtains the following expression for the bound-electron pressure

\begin{eqnarray}
P&=&\frac{1}{4\pi}\sum_{n,l}(2l+1)[R_{nl}(r_{ws})]^2\times\nonumber \\
&&\Big[\frac{D_{nl}^2+2D_{nl}-l(l+1)}{r_{ws}^2}+2[\epsilon_{nl}-v(r_{ws})]\Big],
\end{eqnarray}

which is exactly formula (\ref{expresso}).

\end{document}